\documentclass[prl,twocolumn,nofootinbib,superscriptaddress,preprintnumbers,amsmath,amssymb]{revtex4}
\usepackage{graphicx}
\usepackage{color}
\usepackage{amsbsy}
\usepackage{dcolumn}
\usepackage{bm}
\usepackage{upgreek}
\usepackage{epstopdf}
\usepackage[normalem]{ulem}
\begin{document}
\title{Intrinsic Ferromagnetism in the Diluted Magnetic
Semiconductor Co:TiO$_2$ }

\author{H. Saadaoui}

\affiliation{Laboratory for Muon Spin Spectroscopy,
Paul Scherrer Institute, 5232 Villigen, Switzerland}

\author{X. Luo}
\affiliation{School of Materials Science and Engineering,
University of New South Wales, Kensington, New South Wales 2052, Australia}

\author{Z. Salman}
\affiliation{Laboratory for Muon Spin Spectroscopy,
Paul Scherrer Institute, 5232 Villigen, Switzerland}

\author{X. Y. Cui}
\affiliation{School of Aerospace, Mechanical and Mechatronic
Engineering, The University of Sydney, Sydney, New South Wales 2006,
Australia}

\author{N. N. Bao}
\affiliation{Department of Materials Science and Engineering,
National University of Singapore, 119260, Singapore}

\author{P. Bao}
\affiliation{School of Physics, The University of Sydney, Sydney,
New South Wales 2006, Australia}

\author{R. K. Zheng}
\affiliation{School of Physics, The University of Sydney,
Sydney, New South Wales 2006, Australia}

\author{L. T. Tseng}
\affiliation{School of Materials Science and Engineering,
University of New South Wales, Kensington, New South Wales 2052, Australia}

\author{Y.H. Du}
\affiliation{Institute of Chemical and Engineering Science, Agency
for Science, Technology and Research, 1 Pesek Road, Jurong
Island, 627833, Singapore}

\author{T. Prokscha}
\affiliation{Laboratory for Muon Spin Spectroscopy,
Paul Scherrer Institute, 5232 Villigen, Switzerland}

\author{A. Suter}
\affiliation{Laboratory for Muon Spin Spectroscopy,
Paul Scherrer Institute, 5232 Villigen, Switzerland}

\author{T. Liu}
\affiliation{ANKA, Karlsruhe Institute of
Technology, 76344 Eggenstein-Leopoldshafen, Germany}

\author{Y. R.  Wang}
\affiliation{School of Materials Science and Engineering,
University of New South Wales, Kensington, New South Wales 2052, Australia}

\author{S. Li}
\affiliation{School of Materials Science and Engineering,
University of New South Wales, Kensington, New South Wales 2052, Australia}

\author{J. Ding}
\affiliation{Department of Materials Science and Engineering,
National University of Singapore, 119260, Singapore}

\author{S. P. Ringer}
\affiliation{School of Aerospace, Mechanical and Mechatronic
Engineering, The University of Sydney, Sydney, New South Wales 2006,
Australia}

\affiliation{The Australian Institute for Nanoscale Science and
Technology, The University of Sydney, Sydney, New South Wales 2006,
Australia}

\author{E. Morenzoni}
\email[Corresponding E-mail: ]{elvezio.morenzoni@psi.ch}
\affiliation{Laboratory for Muon Spin Spectroscopy,
Paul Scherrer Institute, 5232 Villigen, Switzerland}

\author{J. B. Yi}
\email[Corresponding E-mail: ]{jiabao.yi@unsw.edu.au}
\affiliation{School of Materials Science and Engineering,
University of New South Wales, Kensington, New South Wales 2052, Australia}

\date{\today}
\newcommand{\LSCO}{La$_{2-x}$Sr$_x$CuO$_{4-\delta}$}
\newcommand{\YBCO}{YBa$_{2}$Cu$_{3}$O$_{7-\delta}$}
\newcommand{\BISCO}{Bi$_2$Sr$_2$Cu$_2$O$_{8+\delta}$}
\newcommand{\STO}{SrTiO$_{3}$}
\newcommand{\SRO}{Sr$_2$RuO$_{4}$}
\newcommand{\PCCO}{Pr$_{2-x}$Ce$_x$CuO$_{4-\delta}$}
\newcommand{\CO}{$\rm CuO_2$}
\newcommand{\Li}{${}^8$Li$^+$}
\newcommand{\Lip}{${}^8$Li$^+$}
\newcommand{\NbSe}{NbSe$_2$}
\newcommand{\CTO}{Co$_{0.05}$Ti$_{0.95}$O$_{2-\delta}$}
\newcommand{\CTOx}{Co$_{x}$Ti$_{1-x}$O$_{2-\delta}$}
\newcommand{\cto}[2]{Co$_{#1}$Ti$_{#2}$O$_{2-\delta}$}
\newcommand{\CoTiO}[3]{Co$_{#1}$Ti$_{#2}$O$_{#3}$}
\newcommand{\pr}[1]{10$^{-#1}$}
\newcommand{\TiO}{TiO$_{2-\delta}$}
\newcommand{\msr}{$\mu$SR}
\newcommand{\lem}{LE-$\mu$SR}
\newcommand{\bnmr}{$\beta$-NMR}
\newcommand{\spct}{superconductivity}
\newcommand{\Spct}{Superconductivity}
\newcommand{\spc}{superconductor}
\newcommand{\Spc}{Superconductor}
\newcommand{\etal}{{\it et al.}}
\newcommand{\Htsc}{High-$T_C$ superconductors}
\newcommand{\htsc}{high-$T_C$ superconductors}
\newcommand{\ie}{{\it i.e.}}
\newcommand{\Tc}{$T_c$}

\newcommand{\PRL}[3]{Phys. Rev. Lett. {\bf #1}, {#2} ({#3})}
\newcommand{\APL}[3]{Appl. Phys. Lett. {\bf #1}, {#2} ({#3})}
\newcommand{\PRB}[3]{Phys. Rev. B {\bf {#1}}, {#2} ({#3})}
\newcommand{\PB}[3]{Physica B {\bf {#1}}, {#2} ({#3})}
\newcommand{\PC}[3]{Physica C {\bf {#1}}, {#2} ({#3})}
\newcommand{\Nt}[3]{Nature {\bf {#1}}, {#2} ({#3})}
\newcommand{\Sc}[3]{Science {\bf {#1}}, {#2} ({#3})}
\newcommand{\RMP}[3]{Rev. Mod. Phys. {\bf {#1}}, {#2} ({#3})}
\newcommand{\JPSJ}[3]{J. Phys. Soc. Jap. {\bf #1}, {#2} ({#3})}
\newcommand{\RPP}[3]{Rep. Prog. Phys. {\bf {#1}}, {#2} ({#3})}
\newcommand{\ibid}[3]{{\it ibid}. {\bf {#1}}, {#2} ({#3})}
\newcommand{\mycite}[6]{\bibitem{#1}#2 \etal, #3 {\bf {#4}}, {#5} ({#6}).}
\newcommand{\equ}[2]{\begin{equation}\label{#1}{#2}\end{equation}}
\newcommand{\meq}[2]{\begin{eqnarray}\label{#1}{#2}\end{eqnarray}}
\newcommand{\figs}[2]{Fig. \ref{#1}-(#2)}
\newcommand{\fig}[1]{Fig. \ref{#1}}
\newcommand{\xtoy}[2]{#1 $\times$ 10$^{#2}$}

\newcommand{\tba}{({\bf To be added})}
\newcommand{\jb}{{\bf JB}}
\newcommand{\celsius}{${}^{\circ}$C}
\newcommand{\perc}[1]{{#1} $\%$}

\begin{abstract}
Here we present a study of magnetism in \CTO\ anatase films grown by
pulsed laser deposition under a variety of oxygen partial pressures
and deposition rates. Energy-dispersive spectrometry and transition electron microscopy
analyses indicate that a
high deposition rate leads to a homogeneous microstructure, while
very low rate or postannealing results in cobalt clustering. Depth
resolved low-energy muon spin rotation experiments show that films grown at a low
oxygen partial pressure ($\approx 10^{-6}$ torr) with a uniform
structure are fully magnetic, indicating intrinsic ferromagnetism.
First principles calculations identify the beneficial role of low
oxygen partial pressure in the realization of uniform
carrier-mediated ferromagnetism. This work demonstrates that
Co:TiO$_2$ is an intrinsic diluted magnetic semiconductor.
\end{abstract}
\maketitle
Diluted magnetic semiconductors (DMSs) have received intensive attention
because of their potential applications in
spintronics devices, which are based on novel
concepts utilizing both the charge and spin of the
electron. One challenging hurdle is achieving intrinsic ferromagnetism
at room temperature. Oxide based DMSs are the most promising
candidates owing to their possible high Curie
temperature \cite{DietlSc00,MatsumotoSc01,PhilipNM06,YamadaSc11,PanPRL07,YiPRL10},
compared to the III-V compound based DMSs, such as
(Ga,Mn)As \cite{Ohnosci98,OhnoNt00}. However, reproducibility
problems of oxide DMSs have been reported
in several cases, where ferromagnetism is only apparent
in samples prepared in a poor oxygen environment.
Magnetic clusters and secondary phases have been shown
to lead to extrinsic ferromagnetism in these systems
\cite{NeyPRL08,KasparNJP08, KimPRL03}.
Therefore, understanding the origin of ferromagnetism in oxide
DMSs is still an object of intense debate  \cite{DietlRMP14,Sato2010}.

One of the most prominent systems is Co-doped TiO$_2$, the first
reported oxide-based DMS with a Curie temperature $T_{C}$ as
high as 600 K \cite{MatsumotoSc01}. Ferromagnetism can be induced in
paramagnetic \CTOx\ films with an electric field \cite{YamadaSc11},
suggesting that magnetism is carrier mediated. However, Co
clustering in this system has been widely reported to cause room
temperature ferromagnetism ($T_{C}=1388$  K in Co)
\cite{ShindePRL04,KimPRL03}.
Ferromagnetism can be also caused by nucleation of highly Co-enriched anatase nanoparticles \cite{ChambersAPL2003}
The discrepancy of
these results has severely affected the development of DMS
materials. Currently, there is no direct evidence of spatially
uniform ferromagnetism as expected from carrier mediation or
Ruderman-Kittel-Kasuya-Yosida (RKKY) interaction. There is only
evidence of localized structural uniformity by high resolution
transmission electron microscopy (HRTEM)\cite{YamadaSc11}.
Therefore, a detailed investigation of the microscopic magnetic
properties  is required to identify the ideal growth conditions to
achieve structural and intrinsic magnetic uniformity.

In this work, we use low-energy muon spin rotation spectroscopy
(LE-$\mu$SR) \cite{Morenzoni04} as a local magnetic probe to examine
the influence of growth conditions on the magnetism in Co-doped
TiO$_2$ thin films deposited by pulsed laser deposition (PLD). We
find spatially uniform and intrinsic ferromagnetism in \CTO\ films
deposited under an oxygen partial pressure $P_{\rm O_2}$=10$^{-6}$
torr. The structure uniformity of the films is confirmed by x-ray
diffraction (XRD) and HRTEM. The magnetic volume fraction and the
internal field width decrease with increasing oxygen partial
pressure. This work directly proves that an intrinsic DMS system can
be synthesized with both structural and magnetic uniformity.

\begin{figure*}[t]
\includegraphics[width=1\textwidth]{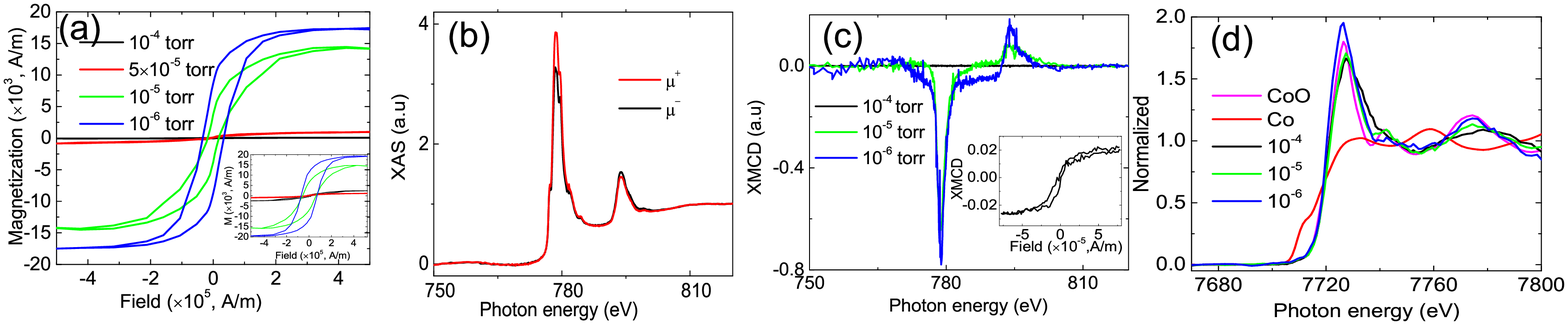}
 \caption{(a) M-H loops  at room temperature of films deposited
 under an oxygen partial pressure of  \pr6, \pr5, 5$\times$\pr5, and \pr4 torr.
 The inset shows the M-H loops at 5 K.
 (b) Co-L edge XAS spectra of the film deposited at \pr5 torr.
 (c) XMCD signal of the films deposited under \pr6, \pr5, and \pr4 torr.
 The inset shows the XMCD  hysteresis of the
 Co-L edge in the sample grown at \pr5 torr.(d) Co-K edge XANES spectra}
 \label{fig-mag}
 \end{figure*}

The films were deposited by PLD in various oxygen partial pressures
with a laser pulse power of 1.8 J/cm$^2$, corresponding to a rate of
10 nm/min. The nominal thickness of the films is approximately 50
nm. The films are epitaxially grown in the anatase structure as
confirmed by XRD. HRTEM imaging shows that films deposited under
$P_{\rm O_2}$=10$^{-5}$ and 10$^{-6}$ torr have a homogeneous
cluster-free microstructure (Fig.S2). Energy dispersive x-ray
spectroscopy (EDX) mapping reveals that Ti, O, and Co atoms are
uniformly distributed in films grown at 10$^{-5}$ and 10$^{-6}$
torr. However, a nonuniform clustered microstructure has been
observed in films deposited under 10$^{-4}$ torr (Fig. S3 in the
Supplemental Material \cite{supplemental}).

\begin{figure}
\includegraphics[width=0.45\columnwidth]{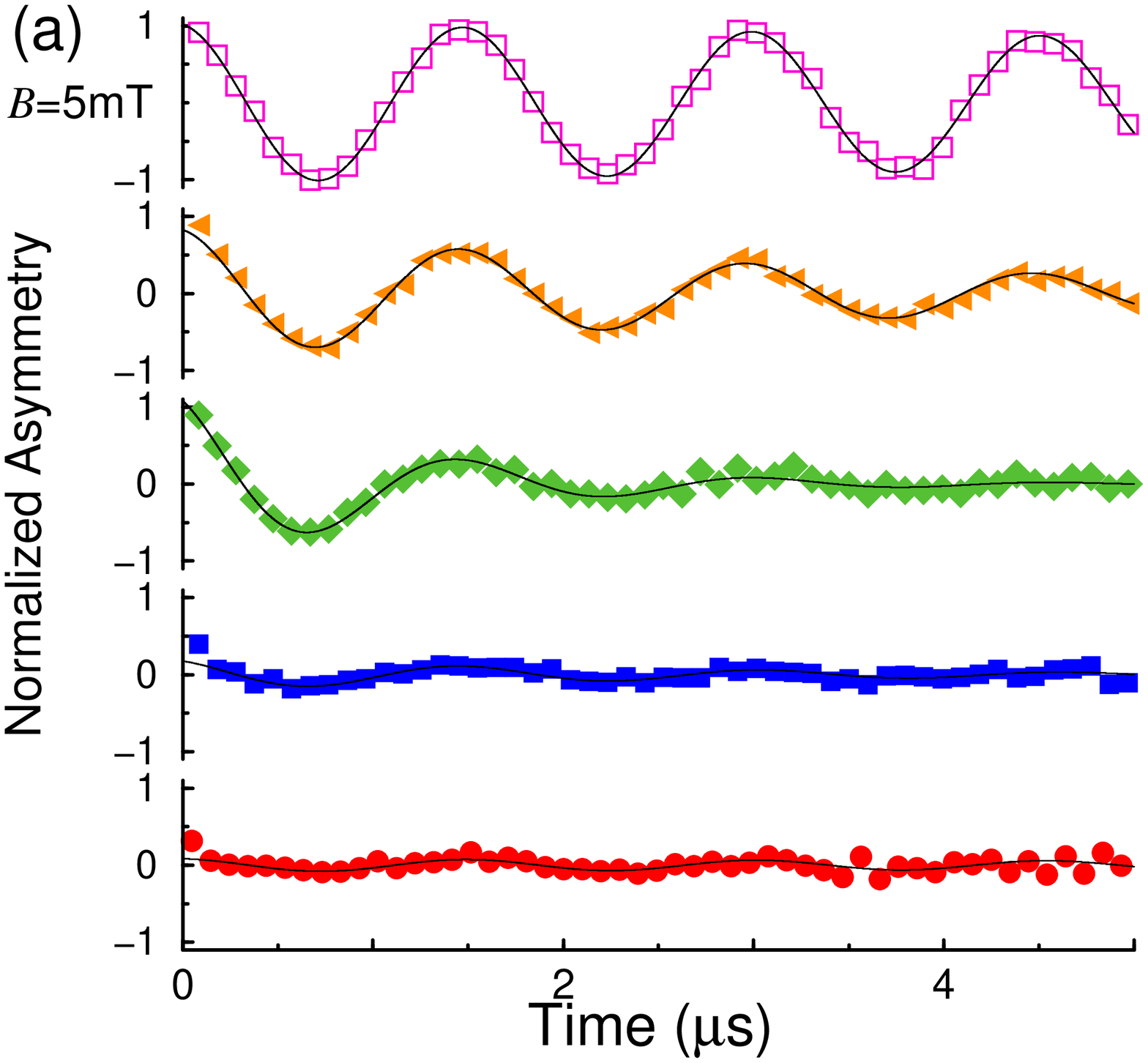}
\includegraphics[width=0.45\columnwidth]{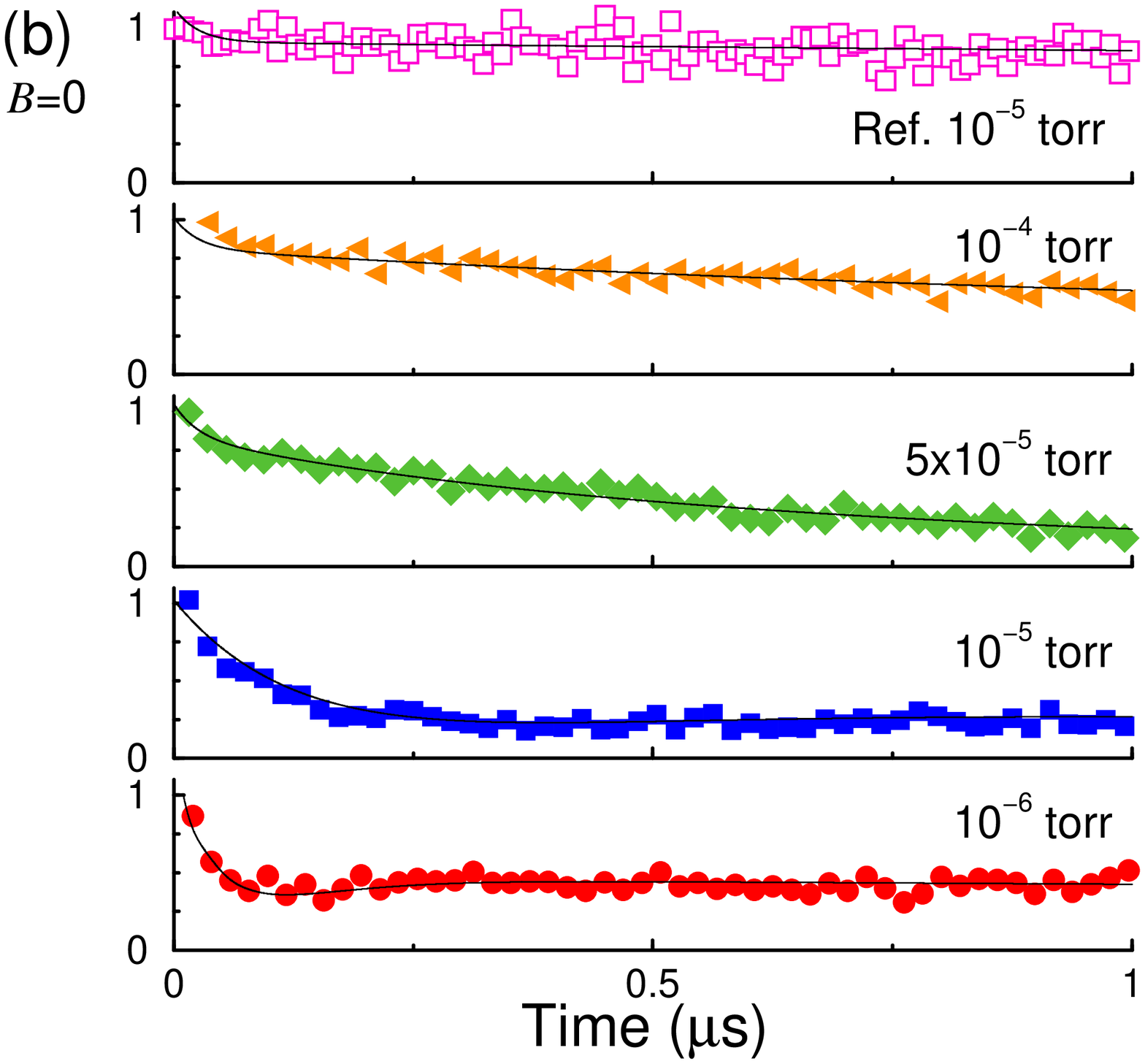}
 \caption{The \lem\ spectra in \CTO \ samples grown at different partial
 oxygen pressures. The spectra are normalized
 to the initial amplitude of the asymmetry of the reference
 sample (undoped TiO$_2$).
 (a) Asymmetry in a transverse field of $B=5$ mT applied parallel
 to the sample surface, measured at 5 K. The initial amplitude at the different oxygen pressures is a measure of the
 paramagnetic fraction as explained in the text.
 (b) Typical asymmetry spectra in zero field.
  All solid lines are the fitted curves described in the supporting material. }
 \label{spectra}
 \end{figure}

Figure  \ref{fig-mag}(a) shows the hysteresis loops of the films
deposited under different oxygen partial pressures. A deposition
pressure of 10$^{-5}$ and 10$^{-6}$ torr leads to magnetic films at
300 K that clearly show hysteresis loops, while growth pressures of
10$^{-4}$ and 5$\times$10$^{-5}$ torr lead to a negligible magnetic
signal. Similar hysteresis loops are seen at 5 K (inset of
Fig.\ref{fig-mag}(a)) with a slight increase of saturation
magnetization.

From the saturation magnetization, we estimate the Co magnetic
moment to be $\sim$1.03 $\mu$$_B$ in the 10$^{-6}$ torr sample and
0.93 $\mu$$_B$ in the 10$^{-5}$ torr sample. Pure TiO$_2$ films
deposited under $P_{\rm O_2}$=10$^{-5}$ torr show no ferromagnetism
at all temperatures, indicating that oxygen vacancies
or Ti$^{3+}$ alone are not the origin of the
ferromagnetism. A similar strong pressure dependence is seen in the
carrier concentration as measured by the Hall effect. The films
deposited under $P_{\rm O_2}$=5$\times$10$^{-5}$, 10$^{-5}$, and
10$^{-6}$ torr have carrier concentrations of 6.4$\times$10$^{17}$,
2.1$\times$10$^{18}$, and 2.0$\times$10$^{19}$ cm$^{-3}$,
respectively, whereas the film deposited under $P_{\rm
O_2}$=10$^{-4}$ torr is insulating.

Figure \ref{fig-mag}(b) shows x-ray absorption spectroscopy (XAS)
spectra at the Co L-edge of the film deposited under $P_{\rm
O_2}$=10$^{-5}$ torr. The multiplet feature of XAS confirms the +2
valence state of Co. Figure \ref{fig-mag}(c) shows the x-ray magnetic circular dichroism  XMCD signal
in three films ($P_{\rm O_2}$=10$^{-4}$, 10$^{-5}$, and 10$^{-6}$
torr) taken at 300 K. A strong signal is observed in the 10$^{-5}$
and 10$^{-6}$ torr samples, whereas no signal is observed in the
film grown at 10$^{-4}$ torr. No magnetic features are observed at
the O and Ti edges, confirming that only Co substitution is the
origin of the observed magnetization in the films. Using sum rules
calculations \cite{TholePRL05}, we estimate the magnetic moments of
Co to be 0.95 and 0.81 $\mu$$_B$, in the 10$^{-6}$ and 10$^{-5}$
torr samples respectively. These estimates are slightly lower than
the values obtained from the magnetization measurements mentioned
above, likely due to surface effects \cite{Singh2012}. The inset
shows the hysteresis of the XMCD signal, where the magnetization of Co
saturates at about 0.2 T, confirming the ferromagnetic phase of the
film. In order to confirm the Co ion valence state, we performed
X-ray absorption near edge spectroscopy (XANES) measurements at the
Co K-edge of the samples and compared the spectra with reference
samples.\cite{Du2015} In all the samples Co show a 2+ state. Because of
lower oxygen partial pressure, Co in the $P_{\rm O_2}$=10$^{-6}$
sample has a slightly lower valence state than in 10$^{-4}$ and
10$^{-5}$. The spectra shown in Fig. \ref{fig-mag}(d) confirm the
absence of metallic Co clusters.

HRTEM and EDX analysis can only give information about the structure
uniformity and chemical composition. On the other hand,
magnetization measurements by a SQUID give only a macroscopic,
sample averaged information of the magnetic state. To determine the
local magnetic properties and the magnetic volume fraction we use
\msr\ as a sensitive magnetic probe \cite{Yaouanc2011}. Fully
polarized muons are implanted in the sample.
The subsequent precession and relaxation of their
spins lead to a temporal evolution of the polarization, which is
easily detectable via the asymmetric muon decay, and can be
used to determine static and dynamic magnetic properties in the
sample. To study the homogeneity of Co-TiO$_2$ thin films
prepared by PLD, we use LE-$\mu$SR which allows a depth dependent
study on a nm scale \cite{Morenzoni04}. This technique has been
successfully applied to a variety of systems such as (Ga,Mn)As DMS
\cite{DunsigerNt10} or oxide heterostructures
\cite{BorisSc11,Morenzoni11,Saadaoui15,SuterPRL}.

\begin{figure}
\includegraphics[width=\columnwidth]{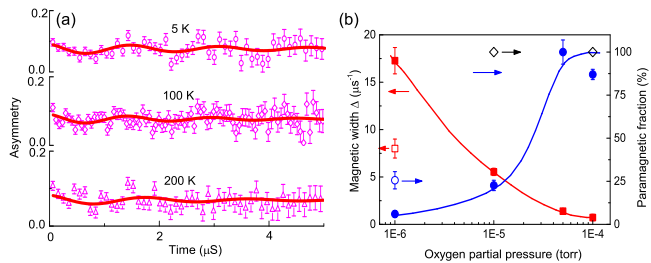}
\caption{(a) TF \lem\ spectra of Co-Ti$O_2$ deposited under
P$O_2$=$10^{-6}$ torr at 5, 100 and 200 K. (b) The width of the
internal magnetic field ($\Delta$) (left $y$-axis) and the
paramagnetic fraction ($F_{pm}$) (right $y$ axis) versus the partial
oxygen pressure during growth. $\Delta$ is found using the ZF
spectra (Fig. \ref{spectra}-(b)). $F_{pm}$ is extracted from the
amplitude of the oscillating part of the asymmetry in TF (Fig.
\ref{spectra}-(a)). The red open square and blue open circle
respectively represent the internal field width and the paramagnetic
fraction in the clustered film. Open black diamonds refer to the
full paramagnetic volume fraction of the nondoped reference
samples. The solid lines are guides to the eye.}

\label{fraction-delta}
\end{figure}

We performed \lem\ measurements in a weak magnetic transverse field
(TF) (5 mT) perpendicular to the substrate surface and the initial
muon spin polarization. Typical spectra are shown in Fig.
\ref{spectra}(a). This measurement allows to determine the magnetic
volume fraction of the sample. In a paramagnetic environment the
local field sensed by the muons is determined by the applied field
and a weakly damped muon spin precession is observed. The undoped
TiO$_2$ sample grown at $P_{\rm O_2}$=10$^{-5}$ torr defines the
reference showing the full asymmetry of a paramagnetic sample
($A_{\rm ref}$). If part of the sample is magnetic muons stopping in
that environment quickly depolarize since field and field width are
much larger than the applied field and do not contribute to the
precession component characteristic of the paramagnetic environment,
whose amplitude $A$ is therefore proportional to the paramagnetic
fraction of the sample. Its dependence on $P_{\rm O_2}$ is therefore
a direct measure of the evolution of the magnetic volume fraction
under different preparation conditions (see the Supplemental Material
for details of the data analysis). The reference spectrum is weakly
damped, while all Co-doped samples are either more damped or have an
initial amplitude lower than the full amplitude of the reference
sample. The dependence of the paramagnetic volume fraction
($F_{pm}=\frac{A}{A_{\rm ref}}\times 100 \%$) on the
partial pressure $P_{\rm O_2}$ is shown in Fig.
\ref{fraction-delta}. The paramagnetic volume fraction increases
with oxygen partial pressure, where paramagnetism and ferromagnetism
coexist in the same sample. The film becomes paramagnetic for
pressures $\gtrapprox$ 5$\times$\pr{5} torr. The sample grown in
10$^{-6}$ torr is almost fully magnetic with a paramagnetic fraction
less than 5$\%$ (within the \lem\ experimental error), due mostly to
the muons landing in the film-free edges of the substrate. Figure 3(a)
shows the temperature dependence of the original TF-$\mu$SR spectra
without any normalization, indicating that there is almost no change
for the spectra at 5, 100 and 200 K, confirming that the
ferromagnetic ordering is not due to the clustering of dopants.

The film deposited under 10$^{-4}$ torr has a $\sim$ 13$\%$
ferromagnetic phase fraction at 5 K. As shown in Fig.
\ref{fig-mag}(a), the film shows a slightly higher saturation
magnetization at 5 K than that of the film deposited at $P_{\rm
O_2}$=5$\times$10$^{-5}$ torr. This residual ferromagnetic phase may
be due to the formation of isolated magnetic polarons in an
insulating state \cite{Coey2005}, which is supported by first
principles density functional theory (DFT) calculations, as
discussed later.

The \lem\ measurements in zero field (ZF) are sensitive to the local
magnetization, since the muon spin relaxation rate $\Delta$ is
proportional to the width of the local field distribution. Typical
asymmetry spectra taken at 5 K with 4 keV muons are shown in Fig.
\ref{spectra}(b). At this energy the muon implantation profile with
a mean stopping depth of 21.2 nm and a rms of 7.1 nm covers the
central part of the film with thickness 50 nm (see Fig. S8 in the
Supplemental Material). The samples grown at low pressures ($P_{\rm
O_2}$=\pr{5} and \pr{6} torr) do not show any spontaneous
precession, as we would expect in the case of long range
ferromagnetic order with well defined wave vector, but a rapid
damping of the muon spin polarization, characteristic of the muons
experiencing a very broad distribution of static fields, due to the
random position of the muons with respect to the magnetic moments.
The signal in the undoped reference sample is slowly relaxing
typical of a nonmagnetic sample. The signal in the film grown at
\pr{4} torr sample shows a fast relaxation at early times with a small
amplitude reflecting a $\approx$ 10\% fraction of clustered
ferromagnetic phases and a dominant slowly relaxing component as
expected in a paramagnetic state, while the sample grown at
5$\times$10$^{-5}$ shows a signal of intermediate relaxation between
the weakly and strongly magnetic samples.

The width of the internal magnetic field $\Delta$ is shown in Fig.
\ref{fraction-delta}. It is weakly dependent on temperature but
decreases strongly with increasing oxygen partial pressure. It is
high at low pressure and approaches the limit of the paramagnetic
reference sample in samples grown at pressures above $5\times$\pr{5}
torr. From the ZF- and TF \lem\ measurements, we conclude that the
film deposited under 10$^{-6}$ torr shows spatially homogeneous
ferromagnetism encompassing the full volume fraction and suggesting
a uniform distribution of Co dopants. This ordering is mediated by
itinerant carriers, since similar films deposited under a high
oxygen partial pressure show weak or no ferromagnetism. Varying the
implantation energy \lem\ allows us also to follow the magnetic
properties as a function of depth on a nanometer scale
\cite{Morenzoni02}. We find the results only weakly dependent on the
implantation depth, further confirming the homogeneous character of
the ferromagnetism.  No evidence of ferromagnetism is observed at
the interface of undoped TiO$_2$ and the LaAlO$_3$ substrate, which
was suggested to be one of the origins
\cite{YoonJMMM07,WeissmannXiv10}.

The \lem\ and structure results rule out Co clustering as an
extrinsic origin of ferromagnetism. We find that the rate of
deposition influences the nature of the magnetic phase. Using HRTEM
and EDX mapping we have observed Co clustering in films deposited
under $P_{\rm O_2}$=10$^{-6}$ torr with a low deposition rate (1
nm/min), or postannealed in vacuum under different oxygen partial
pressures (Supplemental Material). This suggests that the relatively
fast and nonequilibrium deposition at low $P_{\rm O_2}$ may be
impeding the diffusion of oxygen vacancies leading to a uniform
distribution of Co dopants \cite{GrifinPRL05}. In the clustered
10$^{-6}$ film, the XMCD measurement shows a Co moment of $\sim$
0.90 $\mu_B$, slightly lower than  1.03 $\mu_B$ measured in
cluster-free sample. A larger difference is measured by the local
muon probe. The \lem\ measurements show a smaller magnetic fraction
of $\sim$75\% in the clustered film and a field width $\Delta$ that
is about half of that in the strongly magnetic samples (see open
symbols in Fig. \ref{fraction-delta}). The relatively high magnetic
volume fraction indicates that the ferromagnetism in this sample is
not entirely due to Co clustering. With a Co concentration of only 5
$\%$, and taking into account the contribution of stray fields to
the muon signal, we would expect a much smaller magnetic volume
fraction than the observed. Therefore, substitutional Co mainly
contributes to the large fraction of ferromagnetic phase. This
result also confirms the intrinsic nature of relevant ferromagnetic
ordering in Co doped TiO$_2$, no matter whether there are dopant
clusters or not.

To understand the underlying mechanism of the dopant distribution
and the origin of the observed ferromagnetism in Co doped TiO$_2$,
we have performed extensive DFT calculations within the generalized
gradient approximation \cite{Perdew96}, using the Vienna Ab initio
Simulation Package (VASP) code \cite{Kresse1996}, with an on-site
Coulomb repulsion of  U=6.8 $eV$ for Ti 3$d$ orbitals and U=6.5 $eV$
for Co 3$d$ orbitals \cite{Anisimov1991}. First, we calculated the
formation energy of various defects and their complexes in different
charged states \cite{Van2004,Cui2007} as a function of Fermi energy
(from $p$-type to $n$-type growth conditions), as shown in Fig. 4(a).
The interstitial Co ion, Co$_{int}$$^{2+}$, has high formation
energy in the full range of Fermi energies; thus it is unlikely to
play a significant role in mediating the magnetism. A low oxygen
pressure favors the formation of oxygen vacancies (V$_O$) and Ti
interstitial (Ti$_{int}$). Isolated V$_O$ is only stable in the +2
state and Ti$_{int}$ only in the +4 state, both acting as efficient
charge donors, resulting in a $n$-type conductivity. Both V$_O$$^{2+}$
and Ti$_{int}$$^{4+}$ ions are nonmagnetic, indicating that the
magnetic moment is mainly from Co dopants and their complexes, which
is in agreement with Ref. \cite{Weng2004,Yang2003, Kizaki2009} and our XMCD
results. Isolated substitutional Co$_{Ti}$ is stable in the (0) and
(-2) (i.e. Co$^{2+}$ replacing Ti$^{4+}$)charged states.

Interestingly, isolated Co$^0$$_{Ti}$ and Co$^{2-}$$_{Ti}$ induce
similar moments, 0.93 $\mu$$_B$ and 0.99 $\mu$$_B$, respectively,
compared well with our experimental results. A Co$_{Ti}$ dopant and
a nearby V$_O$ may form a Co$_{Ti}$+V$_O$ complex [8], with a (2+/0)
transition level at 1.8 $eV$. (Co$_{Ti}$+V$_O$)$^{2+}$ and
(Co$_{Ti}$+V$_O$)$^0$ ) possess spin moment 1.89 and 0.95 $\mu$$_B$,
respectively. In the the $n$-type limit (low O partial pressure), such
a complex is energetically rather unfavorable, i.e. Co$_{Ti}$$^{2-}$
and V$_O$$^{2+}$ prefer to be separated. The interaction of the
neutral and charged Co dopants and complexes is shown in Fig. 4(b).
It is found that neutral Co$_{Ti}$ dopants prefer to form embedded
clusters as the energy increases monotonically with ion separation.
This is in line with previous DFT calculations [9]. Importantly, the
nearest pair-Co$_{Ti}$ couple ferromagnetically (more stable in
energy by 123 meV over the antiferromagnetic coupling), each
possessing a moment of 0.93 $\mu$$_B$. Three- or four-Co$_{Ti}$
closely-packed clustering structures also couple ferromagnetically,
each Co$_{Ti}$ with a moment of approximately 0.7 $\mu$$_B$. For
both neutral and charged Co$_{Ti}$+V$_O$ complex, the cluster
configuration is energetically favored. However, such clustering
behavior is significantly suppressed when the separation of
complexes is larger than 4 \AA. A further separation actually
lowers the relative total energy, meaning that the driving force
behind clustering is short ranged. Most strikingly, charged
Co$_{Ti}$$^{2-}$ ions prefer to be separated due to the strong
Coulomb force, as the nearest configuration is energetically the
least favorable one. Thus, two factors are identified to effectively
eliminate the clustering of Co dopants; one is the presence of O
vacancy and the other is the $n$-type carrier growth condition. Note
that these two are intrinsically coupled as the O vacancy itself, as
well as the Ti interstitial, is a donor. Thus, the DFT results
highlight the critical role of low oxygen partial pressure in the
fabrication of uniform DMS, as observed in experiments. When the
film is deposited under a low oxygen partial pressure (10$^{-5}$ and
10$^{-6}$ torr), the substitutional Co$_{Ti}$ is uniformly
distributed and long range ferromagnetic ordering is formed via the
mediation of charge carriers. This situation is described in Fig.
S8(c) in the Supplemental Material. When the oxygen partial pressure is relatively high during
film deposition, such as 10$^{-4}$ torr, according to the first
principles calculations, clustered substitutional Co$_{Ti}$$^0$ is
formed since it has low formation energy, resulting in separated
magnetic polarons (note that this is not metallic Co clustering),
which show paramagnetism at room temperature and ferromagnetism at
low temperature. When the oxygen partial pressure is between
10$^{-4}$ and 10$^{-5}$ torr (5$\times$10$^{-5}$ torr),
substitutional Co$_{Ti}$$^{2-}$ is uniformly distributed. However,
in this case the carrier concentration is too low to effectively
mediate the magnetic moment to form ferromagnetic ordering. Hence,
the film is paramagnetic or shows very weak ferromagnetic response.
This is why the film does not have ferromagnetic phase, whereas the
10$^{-4}$ sample has a 13\% ferromagnetic phase.

\begin{figure}
\includegraphics[width=\linewidth]{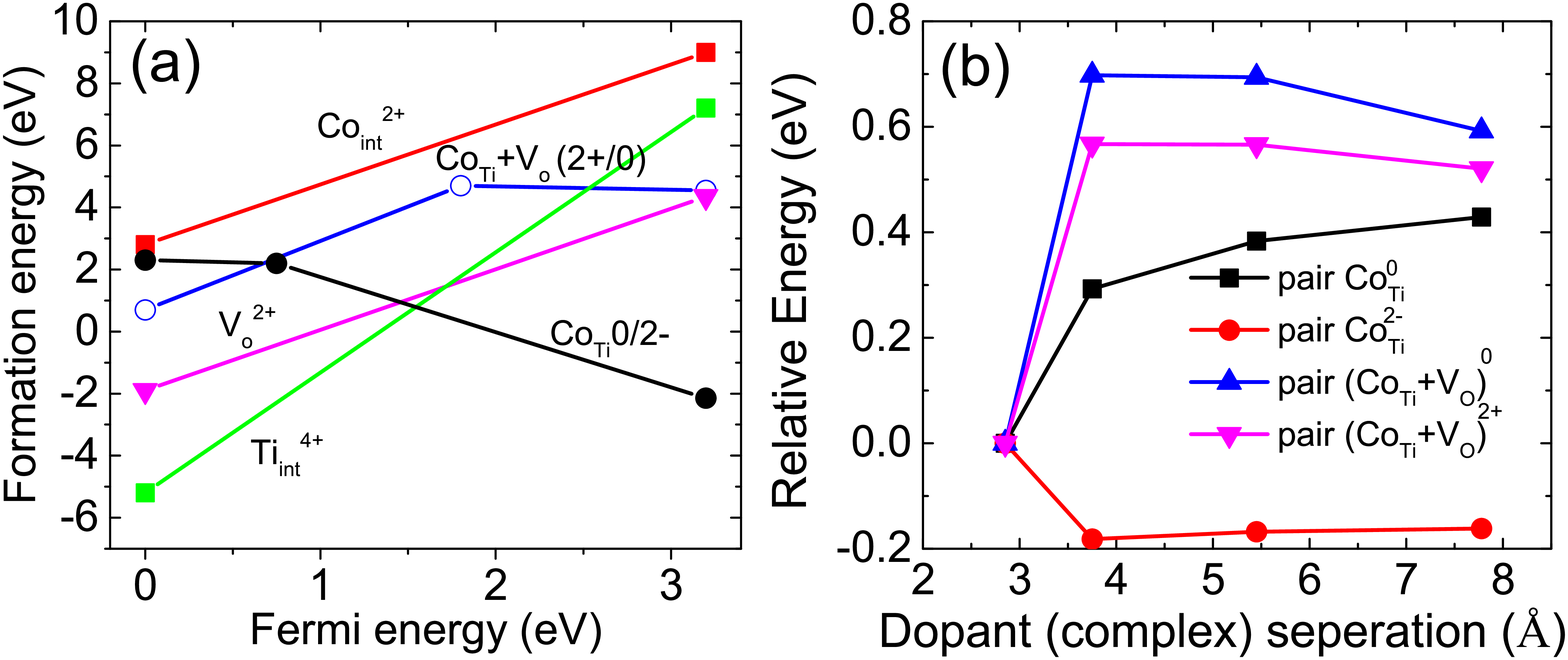}
\caption{(a) Calculated formation energy of the different defects in
neutral and charged states in the Co-doped TiO$_2$ system. (b)
Relative energy of neutral and charged Co$_Ti$ pairs and of
Co$_Ti$ + V$_O$ complexes as a function of separation distance showing that
charge carriers play an important role in the establishment of a
uniform dopant distribution.} \label{Figure4}
\end{figure}

In summary, using depth-resolved \lem, we find that film grown at
\pr6\ torr deposited with relatively high deposition rate is fully
magnetic. The ferromagnetism is intrinsic and controlled by the
oxygen vacancies. Even in the purposely prepared clustered samples,
intrinsic ferromagnetic ordering mainly contributes to a large
fraction of ferromagnetic phase. This work demonstrates that
Co-doped TiO$_2$ prepared under well-controlled parameters, such as
the deposition rate and oxygen partial pressure, has a uniform
structure and intrinsic homogeneous ferromagnetism enforcing its
position as a promising DMS candidate for spintronics devices.

\textbf{Acknowledgments}:
H.S. and E.M. acknowledge the financial support of the MaNEP program.
J.Y. acknowledges the support of
the Australia Research Council discovery project Grants No.
DP110105338 and No. DP140103041.

\end{document}